\documentclass{article} 
\usepackage[sectionbib]{natbib}
\usepackage{array,epsfig,fancyhdr,rotating,float}
\usepackage[]{hyperref}  

\usepackage{sectsty, secdot}
\sectionfont{\fontsize{12}{14pt plus.8pt minus .6pt}\selectfont}
\renewcommand{\theequation}{\thesection\arabic{equation}}
\subsectionfont{\fontsize{12}{14pt plus.8pt minus .6pt}\selectfont}

\usepackage{amsmath}
\numberwithin{equation}{section}
\usepackage{amssymb}
\usepackage{amsfonts}
\usepackage{multirow}
\usepackage{amsthm}

\newtheorem{theorem}{Theorem}

\theoremstyle{definition}

\usepackage{amsmath,amssymb,graphics,epsfig,setspace,multirow,scalerel, comment, url}

\def\cov{{\text{Cov}}} 
\def\be{\mathbb{B}}
\def\re{\mathbb{R}}

\def\var{\text{Var}}

\DeclareMathOperator*{\argmin}{arg\,min}
\renewcommand{\thefootnote}{\fnsymbol{footnote}}

\newcommand{\benu}{\begin{enumerate}}
\newcommand{\eenu}{\end{enumerate}}
\newcommand{\bi}{\begin{itemize}}
\newcommand{\ei}{\end{itemize}}

\usepackage{xcolor}

\usepackage{xcite}
\usepackage{xr}
\usepackage{xr-hyper}
\makeatletter
\newcommand*{\addFileDependency}[1]{
\typeout{(#1)}
\@addtofilelist{#1}
\IfFileExists{#1}{}{\typeout{No file #1.}}
}
\makeatother

\newcommand*{\myexternaldocument}[1]{%
\externaldocument{#1}%
\addFileDependency{#1.tex}%
\addFileDependency{#1.aux}%
}

\myexternaldocument{PICSupp}

\begin{document}


\markright{ \hbox{\footnotesize\rm Statistica Sinica
	}\hfill\\[-13pt]
	\hbox{\footnotesize\rm
	}\hfill }

\markboth{\hfill{\footnotesize\rm TAEHWA CHOI, SANGBUM CHOI AND DIPANKAR BANDYOPADHYAY} \hfill}
{\hfill {\footnotesize\rm PARTIALLY INTERVAL-CENSORED RANK REGRESSION} \hfill}

\renewcommand{\thefootnote}{}
$\ $\par


\fontsize{12}{14pt plus.8pt minus .6pt}\selectfont \vspace{0.8pc}
\centerline{\large\bf RANK-BASED INFERENCE FOR THE }
\vspace{2pt} 
\centerline{\large\bf ACCELERATED FAILURE TIME MODEL WITH}
\vspace{2pt} 
\centerline{\large\bf PARTIALLY INTERVAL-CENSORED DATA}
\vspace{.4cm} 
\centerline{Taehwa Choi, Sangbum Choi and Dipankar Bandyopadhyay
}
\vspace{.4cm} 
\centerline{\it Sungshin Women's University, Korea University and Virginia Commonwealth University}
\vspace{.55cm} \fontsize{9}{11.5pt plus.8pt minus.6pt}\selectfont


\begin{quotation}
\noindent {\it Abstract:}
This paper presents a unified rank-based inferential procedure for fitting the accelerated failure time model to partially interval-censored data. A Gehan-type monotone estimating function is constructed based on the idea of the familiar weighted log-rank test, and an extension to a general class of rank-based estimating functions is suggested. The proposed estimators can be obtained via linear programming and are shown to be consistent and asymptotically normal via standard empirical process theory. Unlike common maximum likelihood-based estimators for partially interval-censored regression models, our approach can directly provide a regression coefficient estimator without involving a complex nonparametric estimation of the underlying residual distribution function. An efficient variance estimation procedure for the regression coefficient estimator is considered. Moreover, we extend the proposed rank-based procedure to the linear regression analysis of multivariate clustered partially interval-censored data. The finite-sample operating characteristics of our approach are examined via simulation studies. Data example from a colorectal cancer study illustrates the practical usefulness of the method. \\ \\ 
\vspace{9pt}
\end{quotation}

\def\thefigure{\arabic{figure}}
\def\thetable{\arabic{table}}

\renewcommand{\theequation}{\thesection.\arabic{equation}}

\fontsize{12}{14pt plus.8pt minus .6pt}\selectfont

\section{Introduction}
\label{sec1}

Interval-censored (IC) responses are frequently encountered in various biomedical settings and health studies, requiring periodic follow-ups or inspections \citep{bogaerts2017}, where the time to an event of interest is only known to fall within a particular interval. Instances of exact failure times observed for certain subjects within the same study leads to partially interval-censored data, which can be categorized into two groups; namely, partly interval-censored \cite[PIC;][]{kim03}, and doubly-censored \cite[DC;][]{ca04}. In addition to a certain number of exact observations (failures/events), DC data are either left-censored or right-censored (e.g., HIV/AIDS studies), whereas PIC data contain additional interval-censoring (e.g., cancer trials). 

The PIC situation commonly arises when studying disease progression, such as the time to progression-free survival in cancer studies, where the nature of periodic examinations for research participants makes it challenging to precisely pinpoint the disease onset or record events, such as time of death, while the study is ongoing. Consequently, some subjects have well-documented information regarding the exact timing of disease onset or occurrence of specific events, while others are interval-censored due to their clinical visit schedules. On the other hand, for DC endpoints, the exact event status is identified only when the measurements are within a specific range. For example, in HIV/AIDS treatment trials, the HIV-1 RNA level is typically used to measure the efficacy of antiretroviral therapy. This measure is generally considered reliable only when the RNA levels fall within a specific range. If the measurements fall outside this range, they are left- or right-censored, leading to DC data. DC and PIC data are reduced to ``Case-1'' and ``Case-2'' interval-censored data, respectively, under absence of exact failure/event times \citep{bogaerts2017}. 

The presence of interval-censoring can introduce significant complexity to the inferential framework. Early research focused on nonparametric estimation of the underlying distribution function via self-consistency equations \citep{tu74}. Many authors delved into exploring the asymptotic behavior and developing computing algorithms for both PIC \citep{zhao08}, and DC \citep{we97} endpoints. For regression, several authors considered a class of asymptotically efficient, yet computationally challenging semiparametric transformation models, which includes proportional hazards (PH) and proportional odds (PO) models, using generalized expectation-maximization (EM) algorithm \citep{pan20}, or direct maximum likelihood (ML) estimation \citep{ch21a}. When the proportional hazards (PH) and proportional odds (PO) assumptions are violated, which is common in cancer trials, a more intuitive alternative is to use an accelerated failure time (AFT) model \citep{choi21}. In this model, the resulting regression estimators can be expressed as a mean ratio, specifically, the ratio of the mean values of the time to the event through a linear model. This approach offers a direct evaluation of the association between event time and covariates.

In this article, we propose a unified rank-based inferential procedure for fitting the AFT regression model to data with PIC and DC endpoints. Let $T_i$ be the failure time, and $X_i$ the $p$-vector of covariates. 
The AFT model specifies that
\begin{equation}\label{aft}
	\log T_i= \beta' X_i+\varepsilon_i,~(i=1,\ldots,n),
\end{equation} 
where $\beta$ is a $p$-vector of unknown regression parameters and $\varepsilon_i$ is a random error with a common but unknown distribution. In contrast to traditional hazard-based approaches, such as the Cox proportional hazards (PH), or proportional odds (PO) models, our AFT specification in (\ref{aft}) directly quantifies the acceleration or deceleration of log-transformed survival times with covariates, thereby offering a more intuitive linear interpretation while bypassing the often restrictive PH/PO assumptions. The classical PH and PO models constructed via the class of linear transformation models \citep{semiparametricchen2002} are amenable to a linear interpretation only under specific parametric choices of the (random) error term, whereas, our AFT framework leaves the error term unspecified (and thus provide flexibility), yet maintaining the linear/direct relationship between the time-to-event response and covariates. 
This feature allows researchers to assess how different factors influence the timing of events, enhancing the practical application of survival models in various settings. 

Various techniques have been developed for estimating regression parameters within the right-censored AFT framework \citep{ji03, cc21}. However, conducting statistical inference with general interval-censored data presents significant challenges, often requiring the simultaneous estimation of both the regression parameters and the underlying residual distribution. To address these challenges, researchers have employed the Buckley-James method \citep{bu79} to handle progressively interval-censored (PIC) data \citep{gao17} and doubly censored (DC) data \citep{choi21}, where the error distribution is approximated using a modified self-consistency algorithm. \cite{li03} introduced a maximum correlation estimating method for AFT models under interval-censoring, but this approach is restricted to scenarios involving only a single covariate. Additional relevant work includes investigations into covariate analysis for Case-2 interval-censored data \citep{tian06}, although these methods are often ad-hoc and challenging to generalize. For Case-1 interval-censored data (commonly referred to as ``current status" data), \cite{gr18} explored nonparametric maximum likelihood (NPML) estimation methods for linear regression, utilizing a kernel-smoothing approach.

In this article, building on the work of \cite{ji03}, we explore a class of weighted log-rank estimating functions for general interval-censored data. Our research shows that the proposed estimators are both consistent and asymptotically normal, a conclusion grounded in standard empirical process theory. A significant advantage of our estimator over existing approaches \citep{gao17, choi21} is that it can directly estimate the regression parameter without needing to nonparametrically estimate the residual distribution function, a process that becomes particularly complex under interval-censoring. To facilitate variance estimation, we introduce an efficient resampling technique \citep{ze08} for estimating covariance matrices for a class of log-rank estimators. This approach eliminates the need to solve estimating equations for resampled data, thereby improving computational efficiency. Additionally, we extend our proposed methods to multivariate scenarios, where cluster sizes may influence event times \citep{kim2010regression}.

The rest of the paper is organized as follows. Section \ref{s:moti} motivates our proposed rank-based approach using a simple data example, leading to the Gehan statistic. Section \ref{s:methods} develops the proposed rank regression framework, along with parameter and standard error estimation for PIC and DC responses. Section \ref{s:sim} studies the finite-sample properties of our proposal using simulated data. The methodology is illustrated in Section \ref{s:apply} through an application to a data example from a colorectal cancer study. 
Finally, Section \ref{s:conc} offers concluding remarks.

\section{Motivation} \label{s:moti}

To illustrate our method, let \(\{1, 2-, 3, 4+, 5\}\) be a simple data of event times. Here, \(\{1, 3, 5\}\) represent exact observations, \(\{2-\}\) is a left-censored observation at 2, and \(\{4+\}\) is a right-censored observation at 4. 
When conducting a linear rank test, our primary goal is to determine the order of two observations within all comparable pairs. For example, although \(\{2-\}\) is left-censored and not fully observable, we can immediately know that \(2- < 3\), \(2- < 4+\), and \(2- < 5\), thereby fixing the order in these cases. However, the ordering between \(1\) and \(2-\) remains indeterminate due to the censored nature of the observations.
Assuming that censoring times are independent of event times, we can perform a Wilcoxon test on this dataset by considering only the comparable pairs, such as \((2-, 3)\), \((2-, 4+)\), and \((2-, 5)\), and determining their ranks within each group.

To be more specific, let us denote observations from two samples by $\{x_1,x_2,\ldots,x_n\}$ and $\{y_1,y_2,\ldots,y_m\}$, which are subject to double-censoring. After pooling the sample of $(n + m)$ observations into a single group, say $\{t_1,t_2,\ldots,t_{n+m}\}$, we can compare each individual with the remaining $n + m - 1$. For comparing the $i$th individual  with the $j$th one, define $G_i=\sum_{j=1}^{m+n}G_{ij}$, where 
\begin{align*}
	G_{ij}=	
	\begin{cases}
	+1, & \text{if}  \ x_i>y_j \  \text{or} \  x_i \geq y_j-  \  \text{or} \    x_i+\geq y_j  \   \text{or} \  x_i+ \geq y_j- \\
	-1, & \text{if} \  x_i<y_j \  \text{or} \ x_i- \leq y_j \  \text{or} \ x_i\leq y_j+   \ \text{or}  \ x_i- \leq  y_j+  \\
	\hfill 0, & \text{otherwise}.\\
\end{cases}
\end{align*}
Thus, for the $i$th individual, $G_i$ is the number of observations that are definitely less than 
$t_i$ minus the number of observations that are definitely greater than $t_i$,
i.e., $G_i=\#\{j:t_i>t_j\}-\#\{j:t_i<t_j\}$. 
Then, the Gehan statistic \citep{ge65} can be defined as 
\begin{equation} \label{wilcox}
	G=\sum_{i=1}^{n+m} G_iI(\text{$i$ in group 1}). 
\end{equation} 

When we consider the null hypothesis that there is no rank difference between two groups, the statistic \( G \) is expected to have a mean of 0. We can estimate its variance as \(\widehat{\var}(G) = \frac{mn}{(m+n)(m+n-1)}\sum_{i=1}^{m+n} G_i^2\). This observation allows us to perform a hypothesis test for comparing two groups within a general interval-censoring framework.
This approach can also be extended to address a linear regression problem involving multiple covariates. By utilizing this idea, we compare the ranks of residuals obtained from the linear model, making it a suitable approach for dealing with general interval-censored  endpoints, which will be discussed in the subsequent sections.

\section{Proposed Methods}\label{s:methods}

\subsection{Partly interval-censored (PIC) rank regression}\label{pic_rank}

We first consider our estimating procedure for PIC data. 
Suppose that there is a sequence of $K$ examination times from periodic clinical visits and let $ 0=W_0 < W_{1} < W_{2}< \cdots < W_{K} < W_{K_i} = \infty$.
We can identify the tightest interval of the examination times $(U,V)$ containing $T$, i.e., $U = \max_{k} \{W_{k} : W_{k} \le T\}$ and $V = \min_{k}\{W_{k}:W_{k} \ge T\}$. 
We use $\Delta$ to represent the censoring indicator, which takes on the value 1 for exact events and 0 for IC cases. When $\Delta_i=1$, we set $U=V=T$. Thus, the PIC dataset is structured as follows: $\{(\Delta_i, \Delta_iT_i, (1-\Delta_i)U_i, (1-\Delta_i)V_i, X_i),~i=1,\ldots,n\}$. Let $\max(a,b)=a\vee b$ and  $\min(a,b)=a\wedge b$. This setup can also be summarized as
$
\{(\tilde U_i,\tilde V_i,\Delta_i, X_i),i=1,\ldots,n\},
$
where 
$\tilde U_{i} = T_i\vee U_i= \Delta_i T_i + (1-\Delta_i )U_i$ and 
$\tilde V_{i} = T_i\wedge V_i = \Delta_i T_i + (1-\Delta_i )V_i$. 
Throughout the paper, we assume that the joint distribution of ($W_1,\ldots,W_K)$ is independent of $T$ given $X$, which implies that the visit processes $\{W_k\}$ do not  provide any further information about the distribution of $T$ given $X$ \citep{zh06}, and that 
the proportion of observing exact observation is non-negligible, i.e., $P(\Delta=1) > 0$.

The rank statistic for PIC data can be built by comparing $(\tilde U_i,\tilde V_i)$ and $(\tilde U_j,\tilde V_j)$ for all pairs of subjects with $i<j$. However, ambiguity arises when either the $i$th subject or the $j$th subject, or both, are interval-censored.
Now, since  $V_i \le U_j$ implies  $ T_i\le T_j $, it suffices to examine 
the following four rank identifiable inequalities:
(i) $T_i\le T_j$
if $(\Delta_i = 1, \Delta_j =1)$,
(ii) $T_i\le U_j$ 
if $ (\Delta_i = 1, \Delta_j =0)$,
(iii) $V_i\le T_j$
if $ (\Delta_i = 0, \Delta_j =1) $, and 
(iv) $V_i\le U_j$
if $ (\Delta_i = 0, \Delta_j =0) $. 
Note that this can be simply accomplished by checking $\tilde V_i \le \tilde U_j$  for all $i<j$ pairs, i.e., whether or not the $i$th  upper bound $\tilde V_i$ is less than the $j$th lower bound $\tilde U_j$. With covariates, let $ u_{i}(\beta) = \log \tilde U_{i} -  \beta' X_i $ and $ v_{i}(\beta) = \log \tilde V_{i} -  \beta' X_i $ denote the observed residuals, corresponding to $\tilde U_i$ and $\tilde V_i$, respectively. 
Also, define $e_i(\beta)=\log T_i-\beta'X_i$ as the true error term that is not directly observable from the data.

By generalizing the Wilcoxon test (\ref{wilcox}) to our regression problem, 
we can formulate the Gehan estimating function as
\begin{equation}\label{eq:efun2}
	S_n(\beta) = n^{-1}\sum_{i=1}^{n}\sum_{j=1}^{n} 
	\eta_{2i}\eta_{1j}( X_i -  X_j)I\{ v_{i}(\beta) \le u_{j}(\beta) \},
\end{equation}
where $\eta_{1i}=I(\tilde U_i>0)=\Delta_i+(1-\Delta_i)I(U_i>0)$ and $\eta_{2i}=I(\tilde V_i<\infty)=\Delta_i+(1-\Delta_i)I(V_i<\infty)$.
Simply speaking, equation \eqref{eq:efun2} takes averages of the covariate differences only when the ordering relationship between $e_i(\beta)$ and $e_j(\beta)$ is ensured for all possible combinations of $(i, j)$.
Indeed, it is clear that $v_i(\beta) \leq u_j(\beta)$ implies  $e_i(\beta) \leq e_j(\beta)$, since the relationship $e_i(\beta) \leq v_i(\beta) \leq u_j(\beta) \leq e_j(\beta)$ holds true in this case.

Nevertheless, as $\beta$ resides within the indicator functions of $S_n(\beta)$ only, the function lacks differentiability, and typically, a unique solution to this estimating function does not exist.
Instead, by recognizing the fact that $S_n(\beta)$ is the negative gradient of the convex objective function
\begin{equation}\label{eq:obj2}
	 L_n(\beta)  = n^{-1}\sum_{i=1}^{n}\sum_{j=1}^{n}
	\eta_{2i}\eta_{1j}\{ v_{i}(\beta) - u_{j}(\beta) \}_{-},
\end{equation}
where $ a_- = |a|I(a\le 0) $, a regression parameter estimate can be obtained by minimizing $L_n(\beta)$ with respect to $\beta$. We define the Gehan estimator as $$\hat\beta=\arg\min_{\beta\in\mathbb{B}} L_n(\beta)$$ The optimizer of this minimization problem may not be unique, but the convexity of $L_n(\beta)$ implies that the set of minimizers is convex \citep{fy94}. The minimization of the equation \eqref{eq:obj2} can be implemented by linear programming, however, efficiency will be lost as larger sample sizes can significantly affect the computational cost. Instead, since $L_n(\beta)$ can be equivalently expressed as 
\begin{equation}\label{eq:optim1}
	\sum_{i=1}^{n}\sum_{j=1}^{n} \eta_{2i}\eta_{1j}  |v_i(\beta) - u_j(\beta)| + 
	\left |M - \beta'\sum_{k=1}^{n}\sum_{l=1}^{n} \eta_{2k}\eta_{1l}( X_l -  X_k) \right |,
\end{equation}
where, $ M >0$ is a sufficiently large number, it can be easily solved by using a standard software, e.g., \texttt{rq()} function in \texttt{R quantreg} package \citep{ko08}.

Furthermore, following \cite{ji03}, we can generalize the Gehan function to the weighted log-rank estimating function as 
\begin{equation}\label{eq:efun3}
	 S_\phi(\beta) = n^{-1}\sum_{i=1}^{n} \eta_{2i}\phi_i(\beta)
	\left[  X_i - 
	\dfrac{\sum_{j=1}^{n}  \eta_{1j}X_jI\{ v_i(\beta) \le u_j(\beta) \}}{\sum_{j=1}^{n}\eta_{1j}I\{ v_i(\beta) \le u_j(\beta) \}}
	\right],
\end{equation}
where $\phi_i(\cdot)$ is a data-dependent nonnegative weight for the $i$th subject.
The choices of $\phi_i(\beta) = \sum_{j=1}\eta_{1j} I\{u_j(\beta)\ge v_i(\beta)\}$  and $\phi_i(\beta) =1$ correspond to the Gehan estimating function \eqref{eq:efun2} and log-rank function, respectively.
The lack of smoothness and monotonicity still imparts computational challenges in solving $S_\phi(\beta)=0$, particularly with multiple covariates.
However, the results of \cite{ji03} imply that there exist a sequence of solutions that is strongly consistent for $\beta$.

Specifically, consider a class of monotone weighted estimating functions
\begin{equation}  \label{eq:iterg2} 
	 S_\phi (\beta, b) =  n^{-1}\sum_{i=1}^{n} \eta_{2i}\eta_{1j}
	 w_i\{b, v_i(b)\} ( X_i -  X_j) I\{ v_i(\beta) \le u_j(\beta) \},
\end{equation}
with the corresponding objective function
\begin{equation}  \label{eq:iterl2}
	 L_\phi (\beta, b) =  n^{-1}\sum_{i=1}^{n}\eta_{2i}\eta_{1j}
	 w_i\{b, v_i(b)\} \{ v_i(\beta) - u_j(\beta) \}_-, 
\end{equation}
where $  w_i(b, t) = \phi_i(b)/\sum_{j=1}^{n} I\{u_j(b)\ge t \} $.
The minimizer of \eqref{eq:iterl2} can be obtained iteratively as 
$\hat\beta_{(k)}= \arg\min_{\beta\in\mathbb{B}} L_\phi(\beta,\hat\beta_{(k-1)}),~k\ge1$,
where, the initial value is set to be the Gehan estimator, i.e., $\hat\beta_{(0)} = \hat\beta$. We define the log-rank estimator by $ \hat\beta_\phi = \lim_{k\to\infty}\hat \beta_{(k)} $. In our experience, these estimators converge in about 5--10 iterative steps, which is not computationally burdensome, and easy to implement.
In Section~\ref{asymp}, we demonstrate that \( \hat \beta_{(k)} \) can be represented as a weighted average of the solutions to \( S_n(\beta) \) and \( S_\phi(\beta) \). This shows that \( \hat \beta_{(k)} \) is consistent for any \( k \ge 1 \) and converges to the solution of \( S_\phi(\beta) \) as \( k\to\infty \).

The methods developed for PIC data can be adapted to DC data with only minor adjustments. For the \(i\)th subject, let \((T_i, L_i, R_i)\) represent the exact event time, left-censoring time, and right-censoring time, respectively. In the context of DC data, we observe the data \(\{(\tilde{T}_i, \delta_i, X_i) \mid i=1,\ldots,n\}\), where \(\tilde{T}_i = L_i \vee (T_i \wedge R_i)\) and \(\delta_i = (\delta_{1i}, \delta_{2i}, \delta_{3i})\) with $\delta_{1i} = I(L_i < T_i \le R_i)$, $\delta_{2i} = I(T_i > R_i)$, and $\delta_{3i} = I(T_i \le L_i)$. Here, \(\tilde{T}_i = T_i\) only if \(\delta_{1i} = 1\); otherwise, it is either right-censored (\(\delta_{2i} = 1\)) or left-censored (\(\delta_{3i} = 1\)). DC data can thus be viewed as a special case of PIC data by defining \((U_i, V_i) = (0, L_i)\) for left-censoring and \((U_i, V_i) = (R_i, \infty)\) for right-censoring. This similarity allows for a unified estimation approach for both PIC and DC data structures. Detailed procedures for estimation under the DC setup are provided in the Supplementary Material S2.


\subsection{Asymptotic properties}
\label{asymp}

In this section, we establish the asymptotic properties of the proposed rank estimators $ \hat\beta $ and $ \hat\beta_\phi$. All technical details (proofs of the Theorems and related Lemmas) are relegated to the Supplementary Material S1. We first impose the following regularity conditions. 

\begin{description}
	\item[](C1) The true value of $\beta$, denoted by $\beta_0$, lies in the interior of a known compact set $\be\subset\re^p $. The covariate $  X $ is uniformly bounded, i.e., $\sup_i\| X_i\|<\infty$ for $ i=1,\ldots,n $. 
	\item[](C2) The residual distribution $F_0\in\mathcal{F}$ is uniformly bounded away from $0$, and has a density with continuous derivative bounded away from $0$ on their support.
    \item[](C3) The distribution of $\Delta$ depends only on the observed data 
    $\{ \Delta, \Delta T, (1-\Delta) U, (1-\Delta)V, X\}$.
    There exists a positive constant $c_0$ such that $P(\Delta=1|X) > c_0$ with probability 1.
    \item[](C4) The joint density of the examination times 
    $(W_1,\ldots,W_K)$ given $\Delta=0$ is continuous and differentiable in their support with respect to some dominating measure.
    There exists a positive constant $\tau_0$ such that 
    $P (\min_{0\le k \le K-1} (W_{k+1}-W_{k}) > \tau_0 |X,K,\Delta=0)=1  $.
\end{description}

Note that, (C1) is a standard assumption in survival analysis implying the compactness of the parameter space with Euclidean norm and boundedness of covariates, while (C2), (C3) and (C4) are essential assumptions that guarantee identifiability of the regression parameters, and strong consistency of their estimators. In particular, (C2) and (C4) state the smoothness conditions for the underlying distribution function. Condition (C3) implies that the IC variables $(U,V)$ do not convey any additional information on the law of $T$ apart from assuming $T$ to be bracketed by $U$ and $V$. This implies that the visit process that generates the $(U,V)$ is independent of $T$  given $X$, and the $U$ and $V$ are constructed from this visit process and from $T$. In addition, the proportion of observing exact time is non-ignorable  to ensure the PIC setup.

\begin{theorem} \label{thm1} \sl
Under conditions (C1)--(C4), the proposed regression estimator $ \hat\beta $ is strongly consistent for $ \beta_0 $, and $ n^{1/2} (\hat\beta-\beta_0) $ converges in distribution to a zero-mean normal distribution with covariance matrix $ \Gamma = A^{-1} \Omega (A^{-1})' $. 
\end{theorem}

\begin{theorem}\label{thm2} \sl
Under conditions (C1)--(C4), for any $ k\ge 1 $, an iterative estimator $ \hat\beta_{(k)} $ is strongly consistent for $ \beta_0 $, and $ n^{1/2} (\hat\beta_{(k)}-\beta_0) $ converges to the  same distribution of 
$ n^{1/2} (\hat\beta_{\phi}-\beta_0) $ as $ k\to\infty $.
\end{theorem}

Theorem \ref{thm1} states the asymptotic behavior of $ \hat\beta $. Here, $A=E\{\partial \psi_\beta/(\partial\beta)|_{\beta=\beta_0}\}$ and $\Omega=E\{\psi_{\beta_0}'\psi_{\beta_0}\}$, where we let $\psi_\beta$ denote
the influence function of $S_n(\beta)$.  The consistency result can be simply proved by using standard arguments of the Glivenko-Cantelli theorem by defining a proper function class composed of indicator functions. For asymptotic normality, we apply the standard asymptotic results of the $Z$-estimator to our estimating functions. Theorem \ref{thm2} shows that the asymptotic behavior  of $ \hat\beta = \lim_{k\to\infty}\hat\beta_{(k)} $ is equivalent to that of $ \hat\beta_\phi $ as the iteration proceeds.

\subsection{Variance estimation}
\label{sec:varest}

A traditional way to estimate standard error for $ \beta $ is to employ resampling-based methods, such as bootstrap and multiplier sampling \citep{ji03,ji06}. However, such multiple resampling-based procedures are usually inefficient, with enhanced computational cost. Instead, we utilize an efficient resampling procedure \citep{ze08} for variance estimation. Note that we can asymptotically write $S_n(\beta)$ as follows:
$$
n^{1/2} S_n(\beta) = n^{1/2} \sum_{i=1}^n \psi_{\beta_0,i}+ n^{1/2} (\beta-\beta_0)A + o_p(1 + n^{1/2} \|\beta-\beta_0\|),
$$ 
where $ A $ is the $p\times p$ invertible matrix (as in Theorem \ref{thm1}), equivalent to the asymptotic slope of $ S_n(\beta) $ at  $\beta_0$. 
To approximate $\Omega$, we use perturbed resampling approach  \citep{ji06a}, which provides more reliable results than standard bootstrapping  under multivariate clustered data. This is partly because bootstrapping under informatively clustered data possibly increases the imbalance in the boostrap data.
Let $R$ be a fixed number of the perturbed resamplings and $S_n^*(\beta)$ be a perturbed estimating function as
$$
S_n^*(\beta) = n^{-1}\sum_{i=1}^n\sum_{j=1}^n Z_i
\eta_{2i}\eta_{1j} (X_i-X_j) I\{v_i(\beta) \le u_j(\beta)\},
$$

where, $Z_i\sim$ i.i.d. Exp(1). Given the sample, $ n^{1/2}S_n^*(\hat\beta) $ asymptotically has a zero-mean normal distribution with the covariance matrix $ \Omega $  \citep{va96}. 
Our variance estimation proceeds as follows:
\begin{description}
	\item[]{\sl Step 1.} Calculate $ R\times p $ matrix $ n^{1/2}S_n^*(\hat\beta) $
	and approximate $\Omega$ by $ \hat \Omega = \widehat\cov\{n^{1/2}S_n^*(\hat\beta)\} $.  
	\item[]{\sl Step 2.} Calculate $ R\times p $ matrix $ n^{1/2} S_n(\hat\beta + n^{-1/2} K_r) $, 
	where $K_r~(r=1,\ldots,R) $  is a $p$-dimensional zero-mean independent random vector, such as $ N(0,1) $.
	\item[]{\sl Step 3.} Regress the $ j $th row of $ n^{1/2} S_n(\hat\beta + n^{-1/2} K_r) $ on $  K_r  $ 
	for $ j=1,\ldots,p $ and $ r=1,\ldots,R $, and 
	let $\hat A$ be the matrix whose $j$th row is the $j$th least squares estimate.
	\item[]{\sl Step 4.} Estimate the covariance matrix of $ n^{1/2}(\hat\beta - \beta_0) $ 
	by $\hat \Gamma = \hat A^{-1} \hat \Omega (\hat A^{-1})'. $
\end{description}

We use a similar inferential method for rank estimation with general weight function. Our numerical studies show that this procedure can produce variance estimators very reliably, achieving nominal coverage probabilities under both PIC and DC settings.

\subsection{Extension to multivariate partly interval-censored data}
\label{sec:cluster}

Next, we extend our methods to handle clustered interval-censored data. This scenario often occurs when subjects are sampled within clusters, leading to correlated failure times among subjects within the same cluster. Suppose that there are $n$ clusters, with the $i$th cluster having $m_i$ members $(m_i\ll n)$. For the $k$th member in the $i$th cluster, let $(T_{ik},U_{ik},V_{ik})$ denote a tuple of  the exact observation, left and right examination times, and 
$\Delta_{ik}=1-I(U_{ik}\le T_{ik}<V_{ik})$ , the censoring indicator, such that $\Delta_{ik}=1$ when $T_{ik}$ is exactly observed. 
The observed multivariate event data under PIC can be represented as 
$\{(\Delta_{ik},\Delta_{ik} T_{ik}, (1-\Delta_{ik})U_{ik}, (1-\Delta_{ik})V_{ik}, X_{ik}),~k=1,\ldots,m_i,i=1,\ldots,n\}$.
Also, define 
$\tilde U_{ik} = T_{ik}\vee U_{ik}= \Delta_{ik} T_{ik} + (1-\Delta_{ik} )U_{ik}$ and 
$\tilde V_{ik} = T_{ik}\wedge V_{ik} = \Delta_{ik} T_{ik} + (1-\Delta_{ik} )V_{ik}$.

We assume the marginal distribution of $T_{ik}$ follows the AFT model
\begin{equation} \label{aft-m}
	\log T_{ik} = \beta' X_{ik} + \varepsilon_{ik},~
	(i=1,\ldots,n; k=1,\ldots,m_i),
\end{equation}
where $(\varepsilon_{i1},\ldots,\varepsilon_{im_i}),~i=1,\ldots,n$ are independent random vectors. Within the $i$th cluster, the error terms,  $\varepsilon_{i1},\ldots,\varepsilon_{im_i}$,  are assumed to be exchangeable with a common marginal distribution $F$. Let 
$u_{ik}(\beta)=\log \tilde U_{ik}-\beta'X_{ik}$
and 
$v_{ik}(\beta)=\log \tilde V_{ik}-\beta'X_{ik}$
denote the observed residuals under model \eqref{aft-m}. To estimate the regression parameter $\beta$, we can solve the generalized log-rank estimating function
\begin{align} \label{multee}
	 S_\phi^\dagger(\beta) = n^{-1}\sum_{i=1}^n\varphi_i\sum_{k=1}^{m_i} \eta_{2ik}\phi_{ik}(\beta)
	\left[ X_{ik} - \dfrac{\sum_{j=1}^n\varphi_j\sum_{l=1}^{m_j}\eta_{1jl} X_{jl} I\{v_{ik}(\beta)\le u_{jl}(\beta) \}  }
	{\sum_{j=1}^n\varphi_j\sum_{l=1}^{m_j}\eta_{1jl} I\{v_{ik}(\beta)\le u_{jl}(\beta) \}}  \right],
\end{align}
where $\eta_{1ik}=\Delta_{ik}+(1-\Delta_{ik})I(U_{ik}>0)$, $\eta_{2ik}=\Delta_{ik}+(1-\Delta_{ik})I(V_{ik}<\infty)$,
and $\varphi_i$ is a known weight to adjust for possible informative cluster sizes (ICS), a setup where the cluster size can be correlated to the survival time of interest  \citep{lam2021marginal}. By convention, we may use $\varphi_i=1$, 
which tends to overweight the large clusters, because each individual contributes equally in the estimating equation. However, when cluster sizes are informative to the outcome of interest, we can incorporate the inverse of cluster sizes as a weight,
for example, $\varphi_i=1/m_i$ or $\varphi_i=1/m_i^{\alpha}$  for some $0\le \alpha\le1$, to relieve the cluster-size effect. This adjustment is also known to  increase statistical efficiency \citep{wa08}. As before, $\phi_{ik}(\beta)$ is a weight function, leading to the Gehan estimator and the log-rank estimator, 
respectively, if $\phi_{ik}(\beta)=\sum_{j=1}^n\varphi_j\sum_{l=1}^{m_j}\eta_{1jl} I\{v_{ik}(\beta)\le u_{jl}(\beta) \}$ and  $\phi_{ik}(\beta)=1$. 

For estimation, we consider the monotone modification of 
$ S_\phi^\dagger(\beta)$ as 
\begin{equation}
	 S_\phi^\dagger(\beta,b) = n^{-1}\sum_{i=1}^n\sum_{k=1}^{m_i}\sum_{j=1}^n\sum_{l=1}^{m_j} \varphi_i\varphi_j  w_{ik}\{b,v_{ik}(b)\}
	\eta_{2ik}\eta_{1jl} (X_{ik}-X_{jl}) I\{v_{ik}(\beta)\le u_{jl}(\beta) \},
\end{equation}
where
$ w_{ik}(b,t)=\phi_{ik}(b)/\sum_{j=1}^n\sum_{l=1}^{m_j}\varphi_j\eta_{2jl}I\{u_{jl}(b)\ge t\}$.
Note that \( S_\phi^\dagger(\beta, b) \) is componentwise monotone in \(\beta\), with the gradient of the convex function given by
\begin{align}\label{mobj}
	 L_\phi^\dagger(\beta,b)= n^{-1}\sum_{i=1}^n\sum_{k=1}^{m_i}\sum_{j=1}^n\sum_{l=1}^{m_j}
	\varphi_i\varphi_j  w_{ik}\{b,v_{ik}(b)\}
	\eta_{2ik}\eta_{1jl} \{v_{ik}(\beta)\le u_{jl}(\beta) \}_-,
\end{align}
which can be minimized again via linear programming or minimizing $\ell_1$-type convex objective function. Let $\hat\beta_{\phi,(0)}^\dagger = \hat\beta^\dagger$. The Gehan estimator $\hat\beta^\dagger$ is easy to implement since $ w_{ik}\equiv1$. The minimization for the log-rank estimator is  carried out iteratively, i.e., $\hat\beta_{\phi,(k)}^\dagger = \argmin_{\beta\in\be}  L_\phi^\dagger(\beta,\hat\beta_{\phi,(k-1)}^\dagger)~(k\ge1) $. 
If the iterative algorithm converges as $k\to\infty$, the limit, say $\hat\beta_\phi^\dagger$,
satisfies the original estimating equation $ S_\phi^\dagger(\beta)$. The consistency and asymptotic normality of \(\hat\beta^\dagger\) and \(\hat\beta_\phi^\dagger\) can be established using similar methods to those outlined for the asymptotic results in Section \ref{asymp}.

To account for the cluster structure in the variance estimation of the marginal model \citep{ji06a,xu23}, we propose using the following perturbed estimating function:
$$
S_n^{\dagger*}(\beta) = n^{-1} \sum_{i=1}^n \sum_{k=1}^{m_i} \sum_{j=1}^n \sum_{l=1}^{m_j} Z_i Z_j \varphi_i \varphi_j \eta_{2ik} \eta_{1jl} (X_{ik} - X_{jl}) I\{v_{ik}(\beta) \le u_{jl}(\beta)\},
$$
where perturbation variables \(Z_i\) and \(Z_j\) are included to reflect the cluster structure \citep{ji06a}. 
To approximate \(\Omega^\dagger\), we use \(\hat\Omega^\dagger = \widehat{\text{Cov}}\{n^{1/2} S_n^{\dagger*}(\hat\beta)\}\). We then update the \(j\)th row of the matrix \(A^\dagger\) with \(\hat{A}^\dagger\), obtained by regressing \(n^{1/2} S_n^\dagger(\hat\beta^\dagger + n^{-1/2} K_r)\) on \(K_r\) (for \(j = 1, \ldots, p\)).
Finally, the covariance matrix of \(n^{1/2} (\hat\beta^\dagger - \beta_0)\) can be estimated by
$
\hat\Gamma^\dagger = \hat{A}^{\dagger-1} \hat\Omega^\dagger (\hat{A}^{\dagger-1})'.$

\section{Simulation Studies} \label{s:sim}

In this section, we present several simulation results using synthetic data to assess the finite-sample performance of our estimates for PIC and DC data, in both univariate and multivariate scenarios. Our method is implemented in the \texttt{R} package \texttt{rankIC}, which includes a detailed vignette. The package is available at the following link: \url{https://github.com/taehwa015/rankIC}.

\subsection{Scenario 1: Univariate data} \label{ss:simuni} 
Here, we generate failure times from the AFT model,
$
\log T= 2+ \beta_1 X_{1}+\beta_2X_{2} + \varepsilon,
$
where $ X_1\sim N(0,1) $ and 
$ X_2  \sim \text{Bernoulli}(0.5)$, and the true regression parameter is set to $(\beta_1,\beta_2)= (1,1)$.
The residual $\varepsilon$ is generated from one of the following three underlying distributions:
(i) standard normal distribution (``$N(0,1)$''),
(ii)  extreme value (EV) distribution (``EV''), and 
(iii) exponential distribution (``Exp(1)''). 
All simulation results are obtained based on  1000 data replications  with sample sizes $ n=200 $ or $ 400 $ and $R=200$ perturbations.

\begin{table}[t]
\caption{Simulation results for PIC data. Table entries are the average bias (Bias), empirical standard error (ESE), asymptotic standard error (ASE), and coverage probability (CP) of the 95\% Wald-type confidence intervals for the parameter estimates obtained from the Gehan and log-rank methods, under $n = 200$ and $400$, censoring rates of $30\%$ (right-censoring: 6\% and interval-censoring: 24\%) and $60\%$ (left-censoring: 1\%, right-censoring: 7\% and interval-censoring: 52\%), and error distributions that follow Normal(0,1) denoted as $N(0,1)$, Extreme Value (EV), and Exponential(1), denoted as Exp(1).}
	\resizebox{\linewidth}{!}{
		\begin{tabular}{|ccccrrrrrrrrr|} 
			\hline
			&&&&\multicolumn{4}{c}{Gehan}&&\multicolumn{4}{c|}{Log-rank}\\ 
			\cline{5-8}\cline{10-13} 
			Cens&Error&$n$&Par&Bias&ESE&ASE&CP&&Bias&ESE&ASE&CP\\
			\hline
			30\%&$N(0,1)$&200&$\beta_1$&--0.005 & 0.076 & 0.074 & 0.942 && 0.007 & 0.081 & 0.079 & 0.941 \\ 
			&&&$\beta_2$&0.000 & 0.146 & 0.146 & 0.946 && 0.003 & 0.161 & 0.156 & 0.943 \\ 
			&&400&$\beta_1$&0.000 & 0.053 & 0.052 & 0.938 && 0.010 & 0.060 & 0.056 & 0.919 \\ 
			&&&$\beta_2$&0.001 & 0.103 & 0.102 & 0.948 && 0.007 & 0.110 & 0.110 & 0.944 \\ 
			&EV&200&$\beta_1$&0.001 & 0.084 & 0.083 & 0.936 && 0.019 & 0.216 & 0.109 & 0.942 \\ 
			&&&$\beta_2$&0.001 & 0.162 & 0.165 & 0.951 && 0.000 & 0.238 & 0.216 & 0.941 \\ 
			&&400&$\beta_1$&0.001 & 0.059 & 0.058 & 0.947 && 0.008 & 0.078 & 0.077 & 0.952 \\ 
			&&&$\beta_2$&0.005 & 0.113 & 0.115 & 0.948 && 0.010 & 0.151 & 0.152 & 0.948 \\ 
			&Exp(1)&200&$\beta_1$&--0.001 & 0.046 & 0.048 & 0.965 && 0.003 & 0.075 & 0.076 & 0.962 \\ 
			&&&$\beta_2$&0.003 & 0.089 & 0.091 & 0.960 && 0.013 & 0.150 & 0.148 & 0.947 \\ 
			&&400&$\beta_1$&--0.001 & 0.031 & 0.032 & 0.951 && 0.005 & 0.053 & 0.053 & 0.952 \\ 
			&&&$\beta_2$&--0.003 & 0.062 & 0.062 & 0.954 && --0.002 & 0.104 & 0.103 & 0.957 \\ \hline
			60\%&$N(0,1)$&200&$\beta_1$&0.000 & 0.074 & 0.075 & 0.948 && 0.023 & 0.085 & 0.083 & 0.938 \\ 
			&&&$\beta_2$&0.003 & 0.155 & 0.148 & 0.935 && 0.016 & 0.173 & 0.160 & 0.929 \\ 
			&&400&$\beta_1$&--0.003 & 0.053 & 0.053 & 0.945 && 0.018 & 0.058 & 0.058 & 0.930 \\ 
			&&&$\beta_2$&0.000 & 0.098 & 0.104 & 0.960 && 0.019 & 0.109 & 0.113 & 0.952 \\ 
			&EV&200&$\beta_1$&--0.004 & 0.084 & 0.083 & 0.946 && 0.015 & 0.112 & 0.109 & 0.941 \\ 
			&&&$\beta_2$&0.005 & 0.163 & 0.164 & 0.945 && 0.014 & 0.219 & 0.213 & 0.939 \\ 
			&&400&$\beta_1$&--0.002 & 0.058 & 0.058 & 0.945 && 0.016 & 0.081 & 0.076 & 0.931 \\ 
			&&&$\beta_2$&--0.008 & 0.117 & 0.115 & 0.944 && 0.008 & 0.159 & 0.150 & 0.936 \\ 
			&Exp(1)&200&$\beta_1$&--0.002 & 0.046 & 0.048 & 0.958 && 0.011 & 0.076 & 0.076 & 0.952 \\ 
			&&&$\beta_2$&0.000 & 0.084 & 0.090 & 0.967 && 0.012 & 0.137 & 0.146 & 0.963 \\ 
			&&400&$\beta_1$&--0.003 & 0.031 & 0.032 & 0.965 && 0.011 & 0.050 & 0.053 & 0.965 \\ 
			&&&$\beta_2$&0.000 & 0.062 & 0.062 & 0.957 && 0.009 & 0.101 & 0.102 & 0.961\\
			\hline 
	\end{tabular}}
	\label{tab1} 
\end{table}

\begin{table}[t!]
\caption{Simulation results for DC data. Table entries are the average bias (Bias), empirical standard error (ESE), asymptotic standard error (ASE), and coverage probability (CP) of the 95\% Wald-type confidence intervals for the parameter estimates obtained from the Gehan and log-rank methods, under $n = 200$ and $400$, ($ \pi_L,\pi_R $) = $(15\%, 15\%)$ and $(30\%, 30\%)$, with $\pi_{L}$ and $\pi_{R}$ denoting proportions of left- and right-censoring, respectively.}
\medskip 
	\resizebox{\linewidth}{!}{\begin{tabular}{|ccccrrrrrrrrr|}
			\hline
			&&&&\multicolumn{4}{c}{Gehan}&&\multicolumn{4}{c|}{Log-rank}\\ 
			\cline{5-8}\cline{10-13} 
			($ \pi_L,\pi_R $)&Error&$n$&Par&Bias&ESE&ASE&CP&&Bias&ESE&ASE&CP \\
			\hline
			(15\%,15\%)&$N(0,1)$&200&$\beta_1$&--0.003 & 0.086 & 0.084 & 0.948 && 0.025 & 0.093 & 0.088 & 0.925 \\ 
			&&&$\beta_2$& 0.003 & 0.168 & 0.163 & 0.934 && 0.005 & 0.180 & 0.173 & 0.930 \\ 
			&&400&$\beta_1$&--0.004 & 0.060 & 0.059 & 0.942 && 0.024 & 0.062 & 0.062 & 0.930 \\ 
			&&&$\beta_2$&--0.002 & 0.111 & 0.115 & 0.958 && 0.001 & 0.119 & 0.122 & 0.963 \\ 
			&EV&200&$\beta_1$&--0.004 & 0.095 & 0.095 & 0.945 && 0.020 & 0.086 & 0.085 & 0.931 \\ 
			&&&$\beta_2$&--0.019 & 0.178 & 0.183 & 0.953 && --0.012 & 0.163 & 0.168 & 0.953 \\ 
			&&400&$\beta_1$&--0.005 & 0.064 & 0.066 & 0.953 && 0.019 & 0.059 & 0.060 & 0.944 \\ 
			&&&$\beta_2$&--0.011 & 0.129 & 0.129 & 0.953 && --0.002 & 0.117 & 0.118 & 0.948 \\ 
			&Exp(1)&200&$\beta_1$&--0.002 & 0.055 & 0.058 & 0.965 && 0.028 & 0.085 & 0.086 & 0.947 \\ 
			&&&$\beta_2$&--0.005 & 0.106 & 0.111 & 0.970 && --0.008 & 0.162 & 0.168 & 0.959 \\ 
			&&400&$\beta_1$&--0.002 & 0.037 & 0.040 & 0.968 && 0.028 & 0.057 & 0.060 & 0.939 \\ 
			&&&$\beta_2$&--0.002 & 0.075 & 0.076 & 0.965 && 0.001 & 0.116 & 0.117 & 0.952 \\ \hline
			(30\%,30\%)&$N(0,1)$&200&$\beta_1$&--0.004 & 0.106 & 0.102 & 0.932 && 0.006 & 0.111 & 0.105 & 0.925 \\ 
			&&&$\beta_2$& 0.003 & 0.206 & 0.205 & 0.938 && --0.026 & 0.211 & 0.209 & 0.939 \\ 
			&&400&$\beta_1$&--0.005 & 0.071 & 0.072 & 0.945 && 0.004 & 0.072 & 0.073 & 0.950 \\ 
			&&&$\beta_2$&--0.001 & 0.142 & 0.144 & 0.959 && --0.028 & 0.148 & 0.148 & 0.942 \\ 
			&EV&200&$\beta_1$&--0.006 & 0.118 & 0.116 & 0.942 && 0.005 & 0.110 & 0.106 & 0.941 \\ 
			&&&$\beta_2$&--0.024 & 0.232 & 0.233 & 0.945 && --0.045 & 0.214 & 0.213 & 0.935 \\ 
			&&400&$\beta_1$&--0.008 & 0.082 & 0.082 & 0.943 && 0.002 & 0.075 & 0.075 & 0.945 \\ 
			&&&$\beta_2$&--0.022 & 0.160 & 0.164 & 0.957 && --0.042 & 0.146 & 0.151 & 0.941 \\ 
			&Exp(1)&200&$\beta_1$&--0.006 & 0.073 & 0.074 & 0.954 && 0.003 & 0.099 & 0.096 & 0.948 \\ 
			&&&$\beta_2$&--0.004 & 0.145 & 0.151 & 0.962 && --0.034 & 0.198 & 0.201 & 0.952 \\ 
			&&400&$\beta_1$&--0.006 & 0.050 & 0.051 & 0.950 && 0.002 & 0.067 & 0.068 & 0.956 \\ 
			&&&$\beta_2$&0.003 & 0.101 & 0.103 & 0.965 && --0.023 & 0.136 & 0.139 & 0.958 \\
			\hline 
	\end{tabular}}   
	\label{tab2}
\end{table}

\begin{table}[t]
\caption{Comparing average Bias (Bias), mean squared error (MSE) and relative efficiency (RE) of the parameter estimates obtained from the Gehan and log-rank methods to the Buckley-James (BJ) type estimators, for data generated under the PIC and DC settings, where $n = 400$, with errors distributed as $N(0,1)$, Extreme Value (EV), and Exp(1), and under censoring rates of 20\% and 40\%. The reported MSEs of the estimators are multiplied by 100. {The Buckley-James (BJ) approach for PIC was implemented via the \cite{gao17}’s method}.} 
\resizebox{\linewidth}{!}{\begin{tabular}{|ccccrrrrrrrrrr|}
			\hline
			&&&&\multicolumn{2}{c}{Buckley-James}&&\multicolumn{3}{c}{Gehan}&&\multicolumn{3}{c|}{Log-rank} \\ \cline{5-6}\cline{8-10}\cline{12-14} 
			Type&Error&Cens&Par&Bias&MSE&&Bias&MSE&RE&&Bias&MSE&RE\\
			\hline
			PIC&$ N(0,1) $&20\%&$\beta_1$&--0.001 & 0.3 && 0.000 & 0.3 & 0.840 && 0.006 & 0.3 & 0.840 \\ 
			&&&$\beta_2$&0.001 & 1.0 && 0.002 & 1.0 & 0.996 && 0.005 & 1.2 & 0.830 \\ 
			&&40\%&$\beta_1$&--0.001 & 0.3 && --0.004 & 0.3 & 0.843 && 0.010 & 0.3 & 0.843 \\ 
			&&&$\beta_2$&0.001 & 1.0 && 0.000 & 1.1 & 0.923 && 0.009 & 1.3 & 0.781 \\ 
			&EV&20\%&$\beta_1$&--0.001 & 0.4 && --0.001 & 0.4 & 1.075 && 0.003 & 0.6 & 0.717 \\ 
			&&&$\beta_2$&0.005 & 1.5 && 0.000 & 1.3 & 1.155 && 0.004 & 2.2 & 0.683 \\ 
			&&40\%&$\beta_1$&0.000 & 0.4 && --0.002 & 0.3 & 1.407 && 0.008 & 0.6 & 0.703 \\ 
			&&&$\beta_2$&0.008 & 1.5 && --0.003 & 1.3 & 1.172 && 0.002 & 2.3 & 0.662 \\ 
			&Exp(1)&20\%&$\beta_1$&0.001 & 0.2 && --0.001 & 0.1 & 2.490 && 0.003 & 0.3 & 0.830 \\ 
			&&&$\beta_2$&0.005 & 1.0 && --0.003 & 0.4 & 2.485 && --0.002 & 1.0 & 0.994 \\ 
			&&40\%&$\beta_1$&0.001 & 0.2 && 0.001 & 0.1 & 2.440 && 0.009 & 0.3 & 0.813 \\ 
			&&&$\beta_2$&0.005 & 1.0 && --0.001 & 0.4 & 2.490 && 0.005 & 1.0 & 0.996 \\
			\hline 
   DC&$ N(0,1) $&20\%&$\beta_1$&0.002 & 0.3 && --0.003 & 0.3 & 0.966 && 0.028 & 0.4 & 0.730 \\ 
			&&&$\beta_2$&0.008 & 1.1 && 0.001 & 1.1 & 1.002 && 0.016 & 1.3 & 0.828 \\ 
			&&40\%&$\beta_1$&0.003 & 0.4 && --0.005 & 0.4 & 0.993 && 0.016 & 0.5 & 0.887 \\ 
			&&&$\beta_2$&0.011 & 1.4 && --0.007 & 1.4 & 1.000 && --0.019 & 1.6 & 0.865 \\ 
			&EV&20\%&$\beta_1$&0.004 & 0.5 && --0.001 & 0.4 & 1.257 && 0.049 & 0.9 & 0.547 \\ 
			&&&$\beta_2$&--0.004 & 1.9 && --0.006 & 1.6 & 1.199 && 0.012 & 2.6 & 0.720 \\ 
			&&40\%&$\beta_1$&0.006 & 0.6 && --0.005 & 0.5 & 1.227 && 0.032 & 0.9 & 0.698 \\ 
			&&&$\beta_2$&--0.003 & 2.3 && --0.018 & 1.9 & 1.173 && --0.040 & 3.1 & 0.728 \\ 
			&Exp(1)&20\%&$\beta_1$&0.001 & 0.3 && --0.001 & 0.1 & 2.428 && 0.033 & 0.4 & 0.704 \\ 
			&&&$\beta_2$&0.004 & 1.2 && 0.000 & 0.5 & 2.434 && 0.016 & 1.3 & 0.926 \\ 
			&&40\%&$\beta_1$&0.002 & 0.4 && --0.004 & 0.2 & 2.339 && 0.019 & 0.4 & 0.982 \\ 
			&&&$\beta_2$&0.011 & 1.4 && --0.002 & 0.6 & 2.211 && --0.014 & 1.4 & 0.982 \\  \hline
	\end{tabular}}
	\label{tab3}
\end{table}

\begin{figure}[h]
	\centering
	\includegraphics[width=0.85\textwidth]{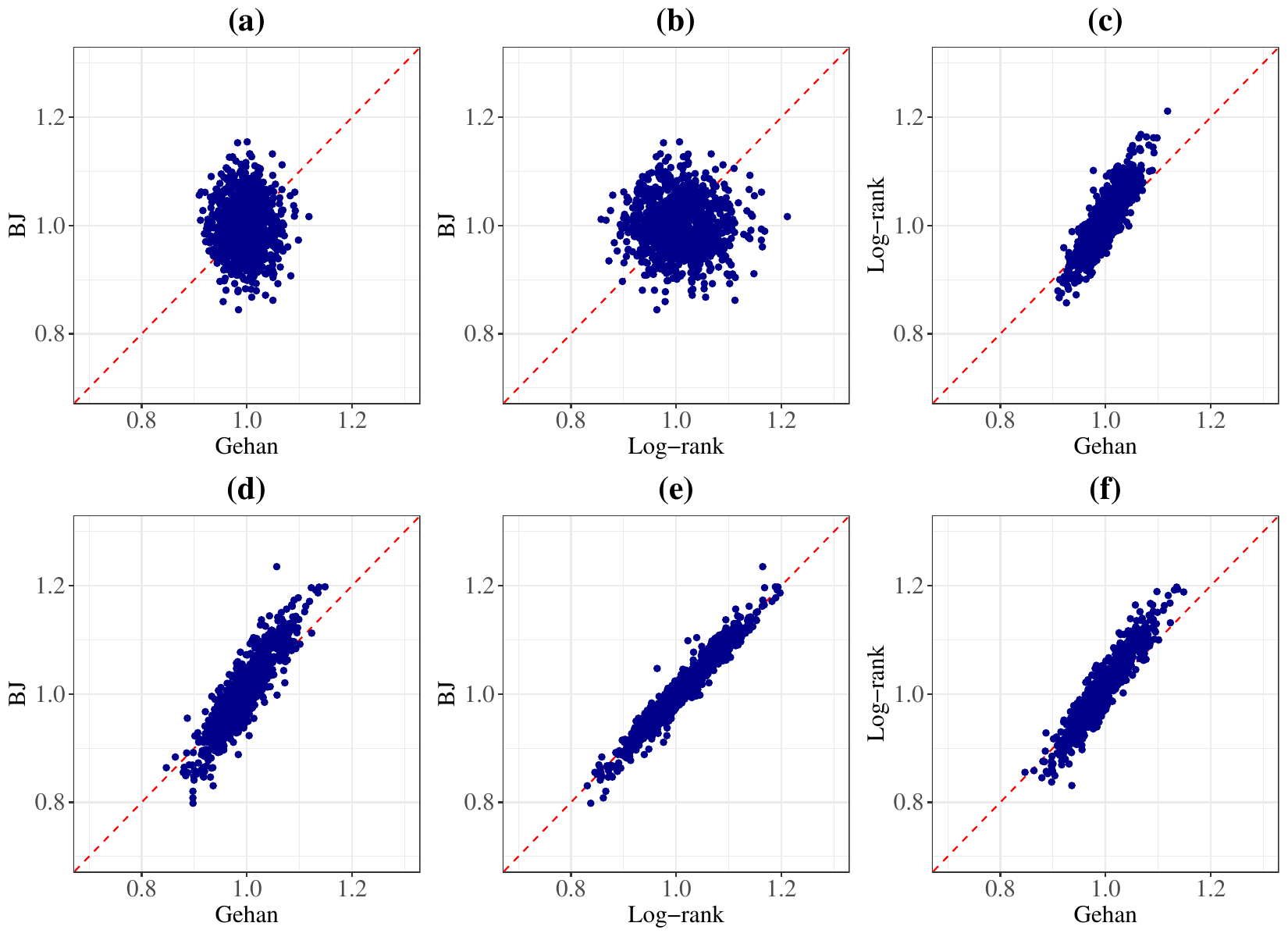}
	\caption{Scatter plots of point estimates of $\beta_{1}$ obtained from the Gehan, log-rank and Buckley-James (BJ) methods, under partly interval-censoring (panels (a)--(c)) and double-censoring (panels (d)--(f)).}
	\label{fig1}
\end{figure}

We first examine the PIC setup, where the proportion of exact observations is given by
$p^* = p_0 - 0.1 \times I(X_{2} = 1), $
with \( p_0 \in (0,1) \) chosen to achieve the desired censoring rate. Conversely, with probability \( 1 - p^* \), the data are subject to interval-censoring. In this scenario, a sequence of random examination times \(\{W_k \mid k=1,\ldots,m\}\) is generated by setting \( W_{k} - W_{k-1} \sim \text{Uniform}(0.1,1) \), ensuring that \( 0 = W_0 < W_1 < \cdots < W_m < \tau \), where \(\tau\) is the maximum follow-up time, set to 100.
Then, we can create the IC data $ (U=W_k,V=W_{k+1}) $ if $ W_{k}<T<W_{k+1} $ for $k\le m-1$. If $T\ge W_m$, we treat $T$ as right-
censored at $W_m$.

We summarize the simulation results comparing the performance of the Gehan and log-rank estimators based on average bias (Bias), empirical standard error (ESE), asymptotic standard error (ASE), and coverage probability of 95\% confidence intervals (CP). Table \ref{tab1} presents the results for sample sizes of \( n = 200 \) and 400 under two scenarios: 30\% censoring (6\% right-censoring and 24\% interval-censoring) and 60\% censoring (1\% left-censoring, 7\% right-censoring, and 52\% interval-censoring) rates. Overall, the proposed estimators are essentially unbiased, and the ASEs closely match their ESEs. Furthermore, the empirical CPs align well with the nominal levels as predicted by normal approximations.  As expected, accuracy improves with larger sample sizes, as evidenced by reductions in MSEs when $n$ increases from 200 to 400.  The Gehan and log-rank estimators show similar performance under \( N(0,1) \) and extreme value (EV) error distributions; however, the Gehan estimator outperforms the log-rank estimator when Exp(1) errors are present. The results corresponding to much larger $n$, i.e., $n=1000$ and $2000$, as presented in Table S1 reveal much smaller Bias, smaller ESEs, and adequate CPs.

In the DC setup, the left-censoring and right-censoring variables \((L, R)\) are simulated as follows: \(\log L \sim (1 - 0.25 X_1) \times \text{Uniform}(-6, c_L)\) and \(\log R \sim \log L + (1 - 0.5 X_2) \times \text{Uniform}(6, c_R)\), respectively. The constants \((c_L, c_R)\) are chosen to achieve the desired left- and right-censoring proportions: \((\pi_L, \pi_R) = (15\%, 15\%)\) and \((30\%, 30\%)\). Table \ref{tab2} summarizes the simulation results for the DC data. 
Similar to the PIC setup, the parameter estimates exhibit small biases that tend to decrease as the sample size \(n\) increases. The standard error estimates accurately reflect the true variability, and the confidence intervals demonstrate appropriate coverage probabilities.

We also compare the statistical efficiency of our rank estimators with those obtained from the Buckley-James (BJ) method under both the PIC \citep{gao17} and DC \citep{choi21} settings. Data for the PIC and DC scenarios are generated as described above, with censoring rates of 20\%and 40\%, and a sample size of \( n = 400 \). Table \ref{tab3} presents the mean bias and mean squared error (MSE) of these estimators, along with the relative efficiency (RE) of our rank estimators (to the BJ estimator), {defined as the ratio of the MSE from the BJ method to the two proposed estimators}. Recall that the BJ estimator, which is essentially a least-square estimator for censored data \citep{lai91}, yields the most efficient maximum likelihood estimate when the residuals are Gaussian. Even under these conditions, our Gehan estimator performs comparably to the BJ estimator. However, for non-normal residual distributions, the Gehan estimator demonstrates superior performance compared to the other methods. Notably, when the error distribution is highly skewed, as in the Exp(1) case, the Gehan estimator achieves MSEs that are nearly 2.5 times smaller than those of the BJ estimator. Additionally, the Gehan estimator yields consistent MSEs across various error distributions, whereas the MSEs of the other methods are more variable, possibly due to their iterative computation processes. This is further illustrated in Figure \ref{fig1}, which shows scatter plots of 1,000 regression estimates of \(\beta_1\), comparing the BJ, Gehan, and log-rank methods under the PIC (upper panel; panels (a)–(c)) and DC (lower panel; panels (d)–(f)) settings. In this scenario, the estimates from the BJ and log-rank methods are quite similar, but the Gehan estimator appears to outperform both. Figure S1 in the Supplementary Material compares the computing times of the three methods. 

\vspace{-0.4in} 

\subsection{Scenario 2: Multivariate data} \label{ss:simmulti}

We further evaluate the proposed methods for clustered PIC and DC settings, where we generate data under the same AFT model (as in Scenario 1) adjusted for the clustered design, specified as
$\log T_{ik} = 2 + \beta_1X_{1ik} + \beta_2X_{2ik} + \nu_i\varepsilon_{ik},
~i=1,\ldots,n;k=1,\ldots,m_i.$ Here, $X_{1ik}\sim N(0,1), X_{2ik}\sim\text{Bernoulli}(0.5)$, $(\beta_1,\beta_2)=(1,1)$, and 
$\varepsilon_{ik}$ follows standard normal distribution. The random effect $\nu_i$ is generated from a gamma distribution with shape parameter $1/\theta$ and scale parameter $\theta$, with $\theta = \{0.5,1\}$. Considering $n=150$, the cluster size is set to $m_i = (r/10) + 2$ (to initiate the ICS scenario),  where $r$ is the $r$th percentile of the distribution of $\nu_i$ satisfying $q_r\le \nu_i <q_{r+10}$, for $r=0,10,\ldots,90$. This configuration yields cluster sizes varying from 2 to 11. The generation of the PIC and DC endpoints follow Section \ref{ss:simuni}. 

Table~\ref{tab4} summarizes the performance of  the Gehan and log-rank estimators via Bias, ESE, ASE and 95\% CP, adjusted for ICS ($\varphi_i = 1/m_i$), or ignoring it ($\varphi_i=1$). The relative efficiencies (RE) of cluster-adjusted method (over unadjusted) are also provided. We observe that although both methods work well, 
the ICS adjusted estimators can achieve higher statistical efficiencies with much lower empirical standard errors, compared to the unadjusted estimators. This indicates that the inverse-size weighting helps reduce standard errors, as well as correct potential ICS issues.

\begin{table}[H]
\caption{Simulation results for multivariate partly interval-censored (PIC) and doubly-censored (DC) data.  Table entries are the average bias (Bias), empirical standard error (ESE), asymptotic standard error (ASE), and coverage probability (CP) of the 95\% Wald-type confidence intervals of the parameter estimates, obtained from the Gehan and log-rank methods for the unadjusted and adjusted analysis, for $\theta = (0.5, 1)$. Also reported are the relative efficiency (RE) of the parameter estimates between the cluster-adjusted and the unadjusted method. }
\medskip 
	\resizebox{\linewidth}{!}{\begin{tabular}{|ccccrrrrcrrrrr|}
			\hline
			&&&&\multicolumn{4}{c}{Unadjusted}&&\multicolumn{4}{c}{Adjusted}& \\ 
			\cline{5-8}\cline{10-13} 
			Type&$\theta$&Method&Par&Bias&ESE&ASE&CP~&&Bias&ESE&ASE&CP~&RE~\\
			\hline
			PIC& 0.5 & Gehan & $\beta_1$&0.001 & 0.051 & 0.051 & 0.954 && 0.001 & 0.042 & 0.041 & 0.942&1.474 \\ 
			&  &  & $\beta_2$&--0.001 & 0.098 & 0.100 & 0.964 && --0.002 & 0.083 & 0.081 & 0.949& 1.393  \\ 
			& & Log-rank & $\beta_1$&0.014 & 0.064 & 0.064 & 0.946 && 0.013 & 0.051 & 0.057 & 0.969&1.549  \\ 
			&  &  & $\beta_2$&0.004 & 0.125 & 0.126 & 0.959 && 0.003 & 0.104 & 0.112 & 0.963 &1.445 \\ 
			& 1 & Gehan & $\beta_1$&0.000 & 0.044 & 0.046 & 0.967 && 0.000 & 0.031 & 0.033 & 0.973 &2.015\\ 
			&  &  & $\beta_2$&--0.005 & 0.090 & 0.089 & 0.950 && --0.004 & 0.064 & 0.063 & 0.957 & 1.976 \\ 
			& & Log-rank & $\beta_1$&0.010 & 0.064 & 0.061 & 0.948 && 0.009 & 0.043 & 0.049 & 0.983&2.174 \\ 
			&  &  & $\beta_2$&--0.003 & 0.126 & 0.120 & 0.944 && 0.001 & 0.085 & 0.094 & 0.968& 2.198 \\
			\hline 
   DC& 0.5 & Gehan & $\beta_1$&--0.004 & 0.058 & 0.059 & 0.948 && --0.003 & 0.048 & 0.050 & 0.955&1.461 \\ 
			&  &  & $\beta_2$&--0.018 & 0.117 & 0.114 & 0.940 && --0.013 & 0.093 & 0.096 & 0.957 &1.589 \\ 
			& & Log-rank & $\beta_1$&0.037 & 0.069 & 0.069 & 0.920 && 0.021 & 0.055 & 0.064 & 0.965 &1.769 \\ 
			&  &  & $\beta_2$&--0.014 & 0.140 & 0.135 & 0.943 && --0.011 & 0.105 & 0.123 & 0.976&1.776  \\ 
			& 1 & Gehan & $\beta_1$&--0.005 & 0.056 & 0.055 & 0.951 && --0.003 & 0.039 & 0.042 & 0.968&2.066   \\ 
			&  &  & $\beta_2$&--0.014 & 0.103 & 0.105 & 0.953 && --0.010 & 0.071 & 0.078 & 0.972 &2.102 \\ 
			& & Log-rank & $\beta_1$&0.039 & 0.069 & 0.068 & 0.919 && 0.018 & 0.044 & 0.056 & 0.986 &2.780 \\ 
			&  &  & $\beta_2$&--0.011 & 0.130 & 0.131 & 0.959 && --0.009 & 0.083 & 0.106 & 0.987&2.442  \\ \hline
	\end{tabular}}    
	\label{tab4}
\end{table}

\section{Application: Metastatic Colorectal Cancer Data}\label{s:apply}

We applied the proposed method to a dataset from a multi-center, randomized, phase III clinical trial on metastatic colorectal cancer \citep{peeters10}. This trial aimed to evaluate the efficacy and safety of panitumumab combined with fluorouracil, leucovorin, and irinotecan (FOLFIRI) compared to FOLFIRI alone, following failure of initial treatment for metastatic colorectal cancer. The covariates considered in our analysis include treatment arm (\(0 =\) FOLFIRI alone, \(1 =\) Panitumumab + FOLFIRI), patient tumor KRAS mutation status (\(0 =\) wild-type, \(1 =\) mutant), scaled patient age at screening, and gender (\(0 =\) male, \(1 =\) female). The primary endpoint is progression-free survival (PFS). 
 \begin{table}[H]
	\caption{ Analysis of colorectal cancer data: Table entries are the estimates, standard errors (SE) and $p$-values of the model parameters obtained from fitting both adjusted (for clustering) and unadjusted AFT models, {using the Gehan, log-rank methods and the Buckley-James method \citep{gao17}}. The estimates corresponding to the unadjusted proportional hazards (PH) and proportional odds (PO) models are also included. }
    \medskip 
    \centering
    \resizebox{\linewidth}{!}{\begin{tabular}{|lrrrcrrrcrrrcrrr|}
			\hline
			& \multicolumn{3}{c}{Treatment} && \multicolumn{3}{c}{Mutation status}&& \multicolumn{3}{c}{Age} && \multicolumn{3}{c|}{Gender}\\
			\cline{2-4}\cline{6-8}\cline{10-12}\cline{14-16}
			Method & 
			\multicolumn{1}{c}{Est}&\multicolumn{1}{c}{SE} &\multicolumn{1}{c}{$p$-value}
			&& \multicolumn{1}{c}{Est}&\multicolumn{1}{c}{SE} &\multicolumn{1}{c}{$p$-value}
			&& \multicolumn{1}{c}{Est}&\multicolumn{1}{c}{SE} &\multicolumn{1}{c}{$p$-value}
			&& \multicolumn{1}{c}{Est}&\multicolumn{1}{c}{SE} &\multicolumn{1}{c|}{$p$-value}\\
			\hline
			\multicolumn{16}{|l|}{Adjusted AFT Model}\\
			Gehan &0.336 &0.102& 0.000 && --0.011 &0.092 & 0.452 && 0.049 &0.058 & 0.196 && --0.262 &0.111 & 0.009 \\ 
			Log-rank &0.256 &0.108 & 0.009 && 0.032 &0.120 & 0.395 && 0.070 &0.054 & 0.098 && --0.324 &0.104 & 0.001 \\ \hline
			\multicolumn{16}{|l|}{Unadjusted AFT Model}\\
			Gehan &0.231 &0.082 & 0.002 && --0.138 &0.070 & 0.025 && 0.057 &0.037 & 0.061 && --0.029 &0.079 & 0.358 \\ 
			Log-rank &0.232 &0.076 & 0.001 && --0.184 &0.083 & 0.013 && 0.096 &0.046 & 0.019 && --0.009 &0.080 & 0.454 \\
            BJ & 0.295 & 0.090 & 0.001 &&--0.190&0.097&0.025&&0.065&0.047&0.086&&--0.027&0.094&0.388  \\\hline
			PH (Unadj.) & --0.206 &0.088 & 0.020 && 0.173 &0.084 & 0.040 && --0.086 &0.046 & 0.058 && 0.006 &0.062 & 0.950\\
            PO (Unadj.) & --0.351 & 0.117 & 0.003 && 0.233 & 0.117&0.047 && --0.101 & 0.067&0.136 && 0.031 & 0.130 & 0.810 \\
			\hline
	\end{tabular}}
	\label{datares2}
\end{table}
Given that tumor response was evaluated every eight weeks during clinic visits, the outcome of interest may be subject to interval-censoring. Setting the baseline assessment at day 0, a patient exhibiting disease progression at the first post-baseline assessment is left-censored, while those showing progression at later assessments are interval-censored. Patients alive without disease progression at the last assessment are right-censored, and exact PFS is observed for patients who died on-study. Among the 855 patients in the dataset, 52 (6.1\%) died on-study, 168 (19.6\%) were left-censored, 329 (38.5\%) were interval-censored, and 306 (35.8\%) were right-censored. Additionally, these patients were distributed across 148 clinic centers, with the number of patients per center ranging from 1 to 23. Previous analyses \citep{pan20} did not account for the potential influence of clinic centers when drawing inferences. Given that the PFS of patients enrolled within the same clinic may be correlated, we propose to account for the effect of the clinic center in our marginal analysis of this data.

\begin{figure}[H]
	\centering
	\includegraphics[width=0.9\textwidth]{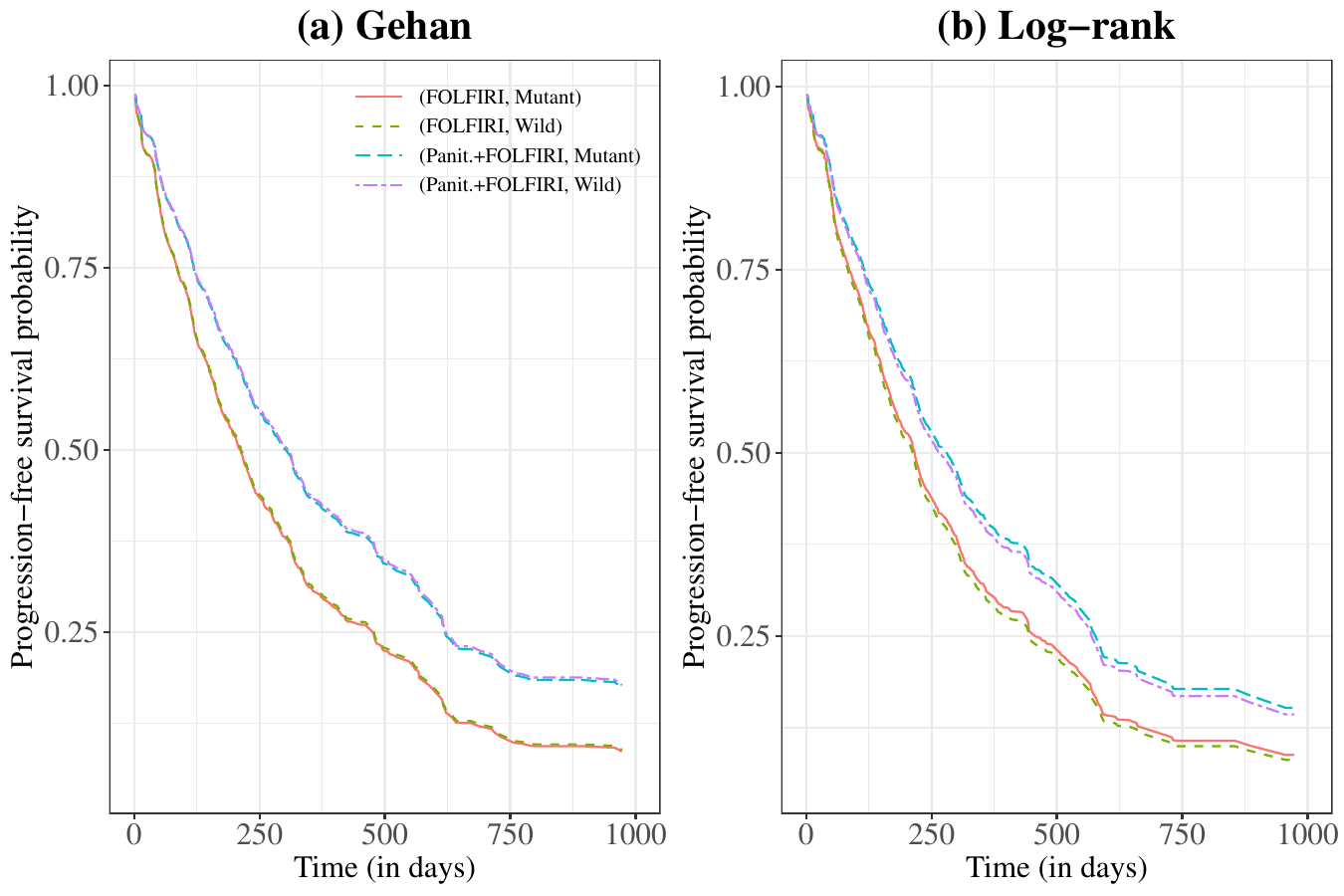}
	\caption{Nonparametric progression-free survival curves corresponding to the 4 groups (combinations of two treatment arms with KRAS mutation status), obtained from the Gehan (Panel a) and log-rank (Panel b) estimators.}
	\label{figres2}
\end{figure}

Table \ref{datares2} presents the mean estimates, standard errors, and associated $p$-values of parameters obtained from fitting both unadjusted and adjusted multivariate AFT models (\ref{aft-m}) using the Gehan and log-rank METHODS {For comparison, the regression results from the Buckley-James method for the AFT model \citep{gao17}, proportional hazards (PH) and the proportional odds (PO) models \citep{an17} are also included}. In all these analyses, the treatment effect is statistically significant, indicating that the panitumumab-combined therapy is more effective in prolonging patients' PFS period. For both the Gehan and log-rank methods, the estimated treatment effect in the adjusted AFT model is higher than in the unadjusted model. On the other hand, KRAS mutation status, which is significant in the unadjusted model, loses its significance in the adjusted model. This phenomenon is also confirmed in Figure \ref{figres2}, which shows that the PFS can increase by approximately 11\% at 1000 days after initial treatment with panitumumab, with no apparent difference between the two mutation types. Here, we use the self-consistent approach \citep{gao17, choi21} for plotting the nonparametric curves with two cluster-size-adjusted rank estimators. The covariate Age remains mostly non-significant, except from the log-rank test in the unadjusted model. On the other hand, Gender, which remains non-significant in the unadjusted model, becomes significant in the adjusted model, implying that the PFS of females is significantly shorter than that of males. \vspace{-0.5in}

\section{Conclusions} \label{s:conc}

In this article, we present a unified rank-based estimation procedure for data with PIC and DC endpoints, extending it to multivariate cases. Most existing work in this area relies on Cox-type models, which often require the simultaneous and complex estimation of both regression and nuisance function parameters. In contrast, our proposed rank-based method directly estimates the regression coefficients without needing to address the residual distribution function, thereby significantly reducing computational complexity. Our simulation studies demonstrate that the proposed rank estimator nearly matches the statistical efficiency of the nonparametric maximum likelihood estimator.

As pointed out by a reviewer, the two rank estimators (Gehan and log-rank) are asymptotically similar, and, in general, we do not claim one to be better than the other. The Gehan estimator is easy to compute, while the log-rank estimator has a very similar form to the Cox estimator. 
If one wants to give variable weights to early events, the Gehan estimator should be used. On the other hand, if one believes that all data points contribute equally, the log-rank estimator is expected to be better.

{An associate editor confirms that the proposed Gehan and log-rank estimators are suboptimal with respect to statistical efficiency, although they are computationally efficient.  To obtain full efficiency, we should utilize the nonparametric function estimator $F$. The optimal weight function should now have the form $\phi^\text{opt}(t) = \lambda'(t)/\lambda(t)$, with the optimal estimating equation \citep{lin13} given as 
$$
S^*_{\phi}(\beta)  = n^{-1}\sum_{i=1}^{n} \phi^\text{opt}(v_i(\beta))\eta_{2i}
	\left[  X_i - 
	\dfrac{\sum_{j=1}^{n}  \eta_{1j}X_jI\{ v_i(\beta) \le u_j(\beta) \}}{\sum_{j=1}^{n}\eta_{1j}I\{ v_i(\beta) \le u_j(\beta) \}}
	\right].
$$
However, this approach would require completely different modeling strategies, and will be pursued in a separate work. 
}

PIC and DC data typically contain a substantial amount of exact failure time observations. In the absence of exact observations, these data types reduce to case-2 and case-1 (or current status) interval-censored data, respectively, and can further generalize to panel count data if some intervals contain more than one count \citep{sun13}. Rank regression analysis for such data is more complex due to the need for intricate theoretical considerations, which are currently being explored by the authors. Additionally, our rank-based estimation procedure does not identify the intercept term in the linear model, necessitating an extra ad-hoc approach to complete the analysis. Consistent and efficient estimation of the intercept term is crucial for survival prediction using the proposed linear model, highlighting a valuable area for future research.

\section*{Supplementary Material}

Proofs of Theorems 1 and 2, details on estimation under the DC setup, and additional simulation studies are relegated to the Supplementary Material.
\par
\section*{Acknowledgments}

The colorectal cancer data was derived based on raw data sets obtained from \url{https://www.projectdatasphere.org/}, which is maintained by Project Data Sphere, LLC. 
The authors acknowledge support from grant (RS-2024-00340298) from the National Research Foundation of S. Korea (T. Choi), grants (2022R1A2C1008514, 2022M3J6A1063595) from the National Research Foundation of S. Korea  and  grant (K2305261) from Korea University (S. Choi), and grants (P20CA252717, P20CA264067, and R21DE031879) from the United States National Institutes of Health (D. Bandyopadhyay).

\par


\bibhang=1.7pc
\bibsep=2pt
\fontsize{9}{14pt plus.8pt minus .6pt}\selectfont
\renewcommand\bibname{\large \bf References}
\expandafter\ifx\csname
natexlab\endcsname\relax\def\natexlab#1{#1}\fi
\expandafter\ifx\csname url\endcsname\relax
\def\url#1{\texttt{#1}}\fi
\expandafter\ifx\csname urlprefix\endcsname\relax\def\urlprefix{URL}\fi

\bibliographystyle{chicago}      
\bibliography{biblist}   

@article{ji06a,
	author = {Jin, Z. and Lin, D. Y. and Ying, Z.},
	journal = {Scandinavian Journal of Statistics},
	number = {1},
	pages = {1--23},
	publisher = {Wiley Online Library},
	title = {Rank regression analysis of multivariate failure time data based on marginal linear 
        models},
	volume = {33},
	year = {2006}
}

@article{semiparametricchen2002,
  title={Semiparametric analysis of transformation models with censored data},
  author={Chen, Kani and Jin, Zhezhen and Ying, Zhiliang},
  journal={Biometrika},
  volume={89},
  number={3},
  pages={659--668},
  year={2002},
  publisher={Oxford University Press}
}

@article{kim2010regression,
  title={Regression analysis of clustered interval-censored data with informative cluster size},
  author={Kim, Yang-Jin},
  journal={Statistics in Medicine},
  volume={29},
  number={28},
  pages={2956--2962},
  year={2010},
  publisher={Wiley Online Library}
}

@article{lam2021marginal,
  title={Marginal analysis of current status data with informative cluster size using a class of semiparametric transformation cure models},
  author={Lam, Kwok Fai and Lee, Chun Yin and Wong, Kin Yau and Bandyopadhyay, Dipankar},
  journal={Statistics in Medicine},
  volume={40},
  number={10},
  pages={2400--2412},
  year={2021},
  publisher={Wiley Online Library}
}

@book{sun13,
	author = {J. Sun and X. Zhao },
	publisher = {New York: Springer},
	title = {Statistical Analysis of Panel Count Data},
	year = {2013}}

@article{cc21,
	author = {T. Choi and S. Choi},
	journal = {Journal of Statistical Computation and Simulation},
	number = {16},
	pages = {3385-3403},
	title = {A fast algorithm for the accelerated failure time model with high-dimensional time-to-event data},
	volume = {91},
	year = {2021}}

@article{wa08,
	author = {Wang, S. and Nan, B. and Zhu, J. and Beer, D. G.},
	journal = {Biometrics},
	number = {1},
	pages = {132--140},
	publisher = {Wiley Online Library},
	title = {Doubly penalized {B}uckley--{J}ames method for survival data with high-dimensional covariates},
	volume = {64},
	year = {2008}}

@article{ch21a,
	author = {Choi, S. and Huang, X.},
	journal = {Communications in Statistics-Theory and Methods},
	number = {9},
	pages = {2188-2200},
	publisher = {Taylor \& Francis},
	title = {Efficient inferences for linear transformation models with doubly censored data},
	volume = {50},
	year = {2021}}

@article{tu74,
	author = {Turnbull, B. W.},
	journal = {Journal of the American Statistical Association},
	number = {345},
	pages = {169-173},
	publisher = {Taylor \& Francis},
	title = {Nonparametric estimation of a survivorship function with doubly censored data},
	volume = {69},
	year = {1974}}

@article{ca04,
	author = {Cai, T. and Cheng, S.},
	journal = {Biometrika},
	number = {2},
	pages = {277-290},
	publisher = {Oxford University Press},
	title = {Semiparametric regression analysis for doubly censored data},
	volume = {91},
	year = {2004}}

@article{bu79,
	author = {Buckley, J. and James, I.},
	journal = {Biometrika},
	number = {3},
	pages = {429-436},
	publisher = {Oxford University Press},
	title = {Linear regression with censored data},
	volume = {66},
	year = {1979}}

@article{ji03,
	author = {Jin, Z. and Lin, D. Y. and Wei, L. J. and Ying, Z.},
	journal = {Biometrika},
	number = {2},
	pages = {341-353},
	publisher = {Oxford University Press},
	title = {Rank-based inference for the accelerated failure time model},
	volume = {90},
	year = {2003}}

@article{ze08,
	author = {Zeng, D. and Lin, D. Y.},
	journal = {Biostatistics},
	number = {2},
	pages = {355-363},
	publisher = {Oxford University Press},
	title = {Efficient resampling methods for nonsmooth estimating functions},
	volume = {9},
	year = {2008}}

@article{ji06,
	author = {Jin, Z. and Lin, D. Y. and Ying, Z.},
	journal = {Biometrika},
	number = {1},
	pages = {147-161},
	publisher = {Oxford University Press},
	title = {On least-squares regression with censored data},
	volume = {93},
	year = {2006}}

@article{li03,
  title={Rank estimation of log-linear regression with interval-censored data},
  author={Li, Linxiong and Pu, Zongwei},
  journal={Lifetime Data Analysis},
  volume={9},
  pages={57-70},
  year={2003},
  publisher={Springer Nature BV}
}

@article{we97,
	author = {Wellner, J. A. and Zhan, Y.},
	journal = {Journal of the American Statistical Association},
	number = {439},
	pages = {945-959},
	publisher = {Taylor \& Francis Group},
	title = {A hybrid algorithm for computation of the nonparametric maximum likelihood estimator from censored data},
	volume = {92},
	year = {1997}}

@article{ge65,
	author = {Gehan, E. A.},
	journal = {Biometrika},
	number = {3/4},
	pages = {650-653},
	title = {A generalized two-sample {W}ilcoxon test for doubly censored data},
	volume = {52},
	year = {1965}}

@article{gao17,
	author = {Gao, F. and Zeng, D. and Lin, D. Y.},
	journal = {Biometrics},
	number = {4},
	pages = {1161-1168},
	publisher = {Wiley Online Library},
	title = {Semiparametric estimation of the accelerated failure time model with partly interval-censored data},
	volume = {73},
	year = {2017}}

@article{zhao08,
	author = {Zhao, X. and Zhao, Q. and Sun, J. and Kim, J. S.},
	journal = {Biometrical Journal},
	number = {3},
	pages = {375-385},
	publisher = {Wiley Online Library},
	title = {Generalized log-rank tests for partly interval-censored failure time data},
	volume = {50},
	year = {2008}}

@article{kim03,
	author = {Kim, J. S.},
	journal = {Journal of the Royal Statistical Society: Series B (Statistical Methodology)},
	number = {2},
	pages = {489-502},
	publisher = {Wiley Online Library},
	title = {Maximum likelihood estimation for the proportional hazards model with partly interval-censored data},
	volume = {65},
	year = {2003}}

@article{ko08,
	author = {Koenker, R.},
	journal = {Journal of Statistical Software},
	number = {6},
	pages = {1-25},
	publisher = {American Statistical Association},
	title = {Censored quantile regression redux},
	volume = {27},
	year = {2008}}

@article{an17,
	author = {Anderson-Bergman, Clifford},
	journal = {Journal of Statistical Software},
	number = {12},
	pages = {1-23},
	title = {icen{R}eg: {R}egression models for interval censored data in {R}},
	volume = {81},
	year = {2017}}

@article{pan20,
	author = {Chun Pan and Bo Cai and Lianming Wang},
	journal = {Statistical Methods in Medical Research},
	pages = {3192-3204},
	title = {A {B}ayesian approach for analyzing partly interval-censored data under the proportional hazards model},
	volume = {29},
    number = {11}, 
	year = {2020}}

@article{tian06,
	author = {L. Tian and T. Cai},
	journal = {Biometrika},
	pages = {329-342},
	title = {On the accelerated failure time model for current status and interval censored data},
	volume = {93},
	year = {2006}}

@article{gr18,
	title={Current status linear regression},
  author={Groeneboom, Piet and Hendrickx, Kim},
  journal={The Annals of Statistics},
  volume={46},
  number={4},
  pages={1415--1444},
  year={2018},
  publisher={JSTOR}}

@article{fy94,
	author = {Fygenson, M. and Ritov, Y.},
	journal = {The Annals of Statistics},
	pages = {732-746},
	title = {Monotone estimating equations for censored data},
	volume = {22},
    number = {2},
	year = {1994}}

@article{peeters10,
  title={Randomized phase {III} study of panitumumab with fluorouracil, leucovorin, and irinotecan ({FOLFIRI}) compared with {FOLFIRI} alone as second-line treatment in patients with metastatic colorectal cancer},
  author={Peeters, M and Price, TJ and Cervantes, A and others},
  year={2010},
  volume = {28},
  number = {31},
  pages = {4706--4713},
  journal={Journal of Clinical Oncology}
}

@article{zh06,
  title={A simple local sensitivity analysis tool for nonignorable coarsening: application to dependent censoring},
  author={Zhang, Jiameng and Heitjan, Daniel F},
  journal={Biometrics},
  volume={62},
  number={4},
  pages={1260--1268},
  year={2006},
  publisher={Wiley Online Library}
}

@article{choi21,
	author = {Choi, Taehwa and Kim, Arlene KH and Choi, Sangbum},
	journal = {Computational Statistics \& Data Analysis.},
	pages = {107306},
	publisher = {Elsevier},
	title = {Semiparametric least-squares regression with doubly-censored data},
	volume = {164},
	year = {2021}}

@book{bogaerts2017,
	author = {Bogaerts, Kris and Kom{\'a}rek, Arno{\v{s}}t and Lesaffre, Emmanuel},
	publisher = {London: Chapman \& Hall/CRC},
	title = {{S}urvival {A}nalysis with {I}nterval {C}ensored {D}ata: {A} {P}ractical {A}pproach with {E}xamples in R, SAS, and BUGS},
	year = {2017}}

@book{va96,
	author = {{van der Vaart}, A.W. and Wellner, J.A.},
	publisher = {New York: Springer},
	title = {{W}eak {C}onvergence and {E}mpirical {P}rocesses: {W}ith {A}pplications to {S}tatistics},
	year = {1996}}

@article{xu23,
  title={Marginal proportional hazards models for multivariate interval-censored data},
  author={Xu, Yangjianchen and Zeng, Donglin and Lin, DY},
  journal={Biometrika},
  volume={110},
  number={3},
  pages={815--830},
  year={2023},
  publisher={Oxford University Press}
}

@article{lai91,
  title={Large sample theory of a modified Buckley-James estimator for regression analysis with censored data},
  author={Lai, Tze Leung and Ying, Zhiliang},
  journal={The Annals of Statistics},
  pages={1370--1402},
  volume={19},
  number={3},
  year={1991},
  publisher={JSTOR}
}

@article{lin13,
  title={Efficient estimation of the censored linear regression model},
  author={Lin, Yuanyuan and Chen, Kani},
  journal={Biometrika},
  volume={100},
  number={2},
  pages={525--530},
  year={2013},
  publisher={Oxford University Press}
}

\vskip .65cm
\noindent
School of Mathematics, Statistics \& Data Science, Sungshin Women’s University, Seoul, South Korea 
\vskip 2pt
\noindent
Center for Data Science, Sungshin Women’s University, Seoul, South Korea
\vskip 2pt
\noindent
E-mail: tchoi@sungshin.ac.kr
\vskip 2pt

\noindent
Department of Statistics, Korea University, Seoul, South Korea
\vskip 2pt
\noindent
E-mail: choisang@korea.ac.kr
\vskip 2pt

\noindent
Department of Biostatistics, Virginia Commonwealth University, Richmond, Virginia, U.S.A.
\vskip 2pt
\noindent
E-mail: dbandyop@vcu.edu

\end{document}